\documentclass[12pt]{iopart}
\usepackage{iopams}
\usepackage{setstack}
\usepackage{graphicx} 
\usepackage{siunitx}
\usepackage{color}
\usepackage{soul}

\begin{document}

\title[Spin detection with a micromechanical trampoline]{Spin detection with a micromechanical trampoline: Towards magnetic resonance microscopy harnessing cavity optomechanics}

\author{R. Fischer$^{1,2}$, D. P. McNally$^1$, C. Reetz$^1$, G. G. T. Assump\c{c}\~{a}o$^1$, T. Knief$^1$, Y. Lin$^{1,\dagger}$, and C. A. Regal$^1$}
\address{$^1$ JILA, National Institute of Standards and Technology and University of Colorado, and
Department of Physics, University of Colorado, Boulder, Colorado 80309, USA}
\address{$^2$ Rafael Ltd, Haifa 31021, Israel}
\address{$^\dagger$ current address: CAS Key Laboratory of Microscale Magnetic Resonance and Department of Modern Physics, University of Science and Technology of China, Hefei 230026, China}
\ead{regal@colorado.edu}

\newcommand{\xzp}{x_{\rm{zp}}} 
\newcommand{\cmt}[1]{{\color{blue}#1}}
\newcommand{\red}[1]{{\color{red}#1}}
\newcommand{\cyan}[1]{{\color{cyan}#1}}
\newcommand{\liny}[1]{{\color{green}#1}}

\begin{abstract}
We explore the prospects and benefits of combining the techniques of cavity optomechanics with efforts to image spins using magnetic resonance force microscopy (MRFM).  In particular, we focus on a common mechanical resonator used in cavity optomechanics -- high-stress stoichiometric silicon nitride (Si$_3$N$_4$) membranes.  We present experimental work with a ‘trampoline’ membrane resonator that has a quality factor above $10^6$ and an order of magnitude lower mass than a comparable standard membrane resonators.  Such high-stress resonators are on a trajectory to reach 0.1 $\rm{aN}/\sqrt{\rm{Hz}}$ force sensitivities at MHz frequencies by using techniques such as soft clamping and phononic-crystal control of acoustic radiation in combination with cryogenic cooling.  We present a demonstration of force-detected electron spin resonance of an ensemble at room temperature using the trampoline resonators functionalized with a magnetic grain.  We discuss prospects for combining such a resonator with an integrated Fabry-Perot cavity readout at cryogenic temperatures, and provide ideas for future impacts of membrane cavity optomechanical devices on MRFM of nuclear spins. 


\end{abstract}


\maketitle

\section{\label{sec:level1}Introduction}
    
The field of cavity optomechanics, in which mechanical motion is well coupled to an optical resonator, has seen rapid progress in recent years, with applications in particular to utilizing and achieving a quantum regime~\cite{aspelmeyer_cavity_2014}. Experimenters have harnessed unique mechanical resonators with both high resonant frequencies, which favor the observation of quantum effects in comparison to thermal scales, and high quality factors that offer environmental isolation. In particular tensioned elements, for example silicon nitride (Si$_3$N$_4$) strings or drums, were found to be well-adapted to cavity optomechanics~\cite{unterreithmeier_damping_2010}, and ultracoherent mechanical tensioned resonators have been enabled by engineering phononic bandgaps and bending profiles ~\cite{yu_phononic_2014,tsaturyan_demonstration_2014,tsaturyan_ultra-coherent_2017,Ghadimi_elastic_2018,teufel_circuit_2011}. A result of this development is a class of mechanical resonators with novel force sensing prospects, thanks to a combination of high force sensitivity, high resonant frequencies, and compatibility with excellent displacement readout.


In this article, we focus on sensing spins in solids with MRFM using  a  membrane  mechanical resonator ~\cite{scozzaro_magnetic_2016} engineered  for high-$Q$ and low mass in a platform compatible with a cavity-optomechanical device. Standard magnetic imaging techniques, utilizing inductive detection schemes, reach sensitivities of $10^{13}$ nuclear spins or $10^{10}$ electron spins. MRFM combines ideas of magnetic resonance imaging (MRI) and scanning probe microscopy in a method that has prospects to significantly reduce the number of spins and the corresponding voxel size that can be detected in magnetic imaging~\cite{sidles_magnetic_1995}. If sensitivity at the single-spin level could be realized, three-dimensional images of molecules such as proteins could be taken at the atomic scale ~\cite{Sushkov_MagneticResonance_2014, ajoy_atomic-scale_2015, lovchinsky_NMR-detection_2016}. MRFM relies upon extreme force sensitivities that have reached the atto-newton level in experiments to date~\cite{mamin_sub-attonewton_2001}. This has enabled measurement of a single electron spin \cite{rugar_single_2004}, or a small ensemble of nuclear spins, corresponding to less than 10 nm resolution \cite{degen_nanoscale_2009, rose_HighResolution_2018}. Achieving the extreme sensitivities required for this demanding imaging technique have been a long-standing challenge~\cite{poggio_force-detected_2010}.  Further, nanoscale magnetometers using nitrogen vacancy (NV) centers in diamond have provided an alternative and rapidly-growing route to nanoscale magnetic imaging~\cite{chernobrod_spin_2005,balasubramanian_nanoscale_2008,degen_nuclear_2008,taylor_pr_2008,maletinsky_robust_2012,pelliccione_two-dimensional_2014,rondin_magnetometry_2014,rugar_proton_2015}.  However, the challenges of achieving a sensitive force sensor at a nanometric scale are orthogonal to realizing a long coherence NV defect close to the diamond surface \cite{ofori-okai_spin_2012, rosskopf_investigation_2014, romach_spectroscopy_2015, kim_decoherece_2015}.

\begin{figure}[t]
\centering
\includegraphics[width=80mm]{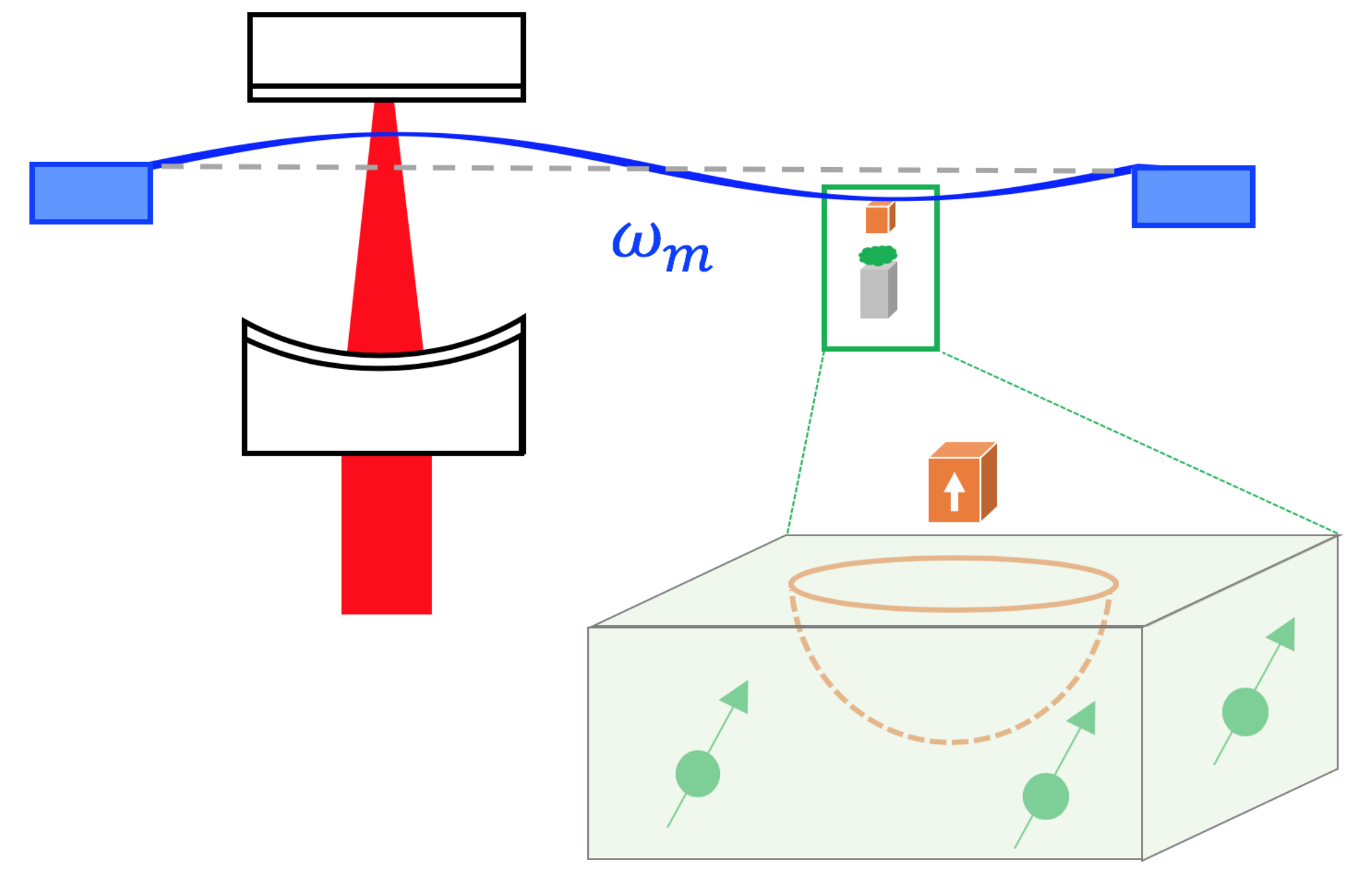}
\caption{Concept of force detected magnetic resonance microscopy with cavity optomechanical coupling. A single mechanical mode is coupled to an optical cavity (red) and to a spin sample (green) at spatially separate locations.  The mechanical mode $(\omega_m)$ could in principle be resonant with the Larmor frequency of nuclear spins in the sample. A magnet (orange) affixed to the resonator creates a large gradient that provides the magnetic coupling and spatial resolution based on magnetic resonance (orange slice). (Note the spin could alternatively be fixed to the resonator, and correspondingly the magnet to an external tip.) An external homogeneous magnetic field $B_0$ dominates the direction and magnitude of the total magnetic field. (See. Fig.~\ref{fig:bdirection} for more detail of bias field and gradient orientations for different detection configurations.)}
\label{fig:concept}
\end{figure}

We identify a number of distinct benefits of a cavity optomechanics platform for force-detected magnetic resonance. In a future fully-integrated system, we envision a concept as shown in Fig.~\ref{fig:concept} in which cavity optomechanical and magnetic coupling are realized simultaneously. Although illustrated with a Fabry-Perot cavity and a silicon nitride (SiN) membrane, the optical integration could take many forms and benefit from a variety of current cavity optomechanics techniques~\cite{eichenfield_optomechanical_2009,schliesser_resolved-sideband_2008,thompson_strong_2008,andrews_bidirectional_2014}.  Electromechanical couplings could also be used if they are designed to tolerate the large magnetic fields required for magnetic resonance~\cite{regal_measuring_2008}.  Cavity optomechanical damping, analogous to active damping commonly used in force sensing ~\cite{rugar_mechanical_1991}, has demonstrated cooling mechanical resonators to their quantum ground state~\cite{chan_laser_2011, teufel_sideband_2011, peterson_laser_2016}. Although this damping does not enhance force sensitivity, deep passive damping using a cavity enables increases in bandwidth combined with an excellent displacement sensitivity.   

 In the context of mechanical devices, engineered SiN membrane resonators offer the promise of potentially record force sensitivities among devices compatible with cavity optomechanics; we project sensitivities below 0.1 $\rm{aN}/\sqrt{\rm{Hz}}$ ~\cite{yu_phononic_2014,tsaturyan_ultra-coherent_2017,Ghadimi_elastic_2018,Yuan_SiNmembrane_2015} in a dilution refrigerator environment. Further, these sensitivities can be achieved at higher frequency of operation - at the MHz scale and likely above. Here one expects to reduce $1/f$ noise encountered due to surface effects, which is a limiting factor in MRFM and has been explored in the context of MHz frequency nanowire sensors~\cite{nichol_nanomechanical_2012}. 

If mechanical frequencies of high-force sensitivity resonators can be pushed to 10 MHz and beyond, direct resonant coupling between the nuclear spins and the mechanical resonator can be achieved at Tesla-scale magnetic fields. Most spin-sensing protocols to date rely on the response of a mechanical resonator to a driven magnetization variation. Resonant coupling could enable distinct readout capabilities and potential access to information on transverse magnetization, as achieved via magnetic induction detection of NMR or electron spin resonance (ESR)~\cite{bargatin_nanomechanical_2003,butler_nanoscale_2010, scozzaro_magnetic_2016}.  Combining strong spin-mechanical resonant coupling with cavity optomechanical coupling to a cold optical bath may enable cooling of the spin ensemble, and corresponding control of the ensemble spin polarization~\cite{magusin_coherent_2000,butler_nanoscale_2010,butler_polarization_2011,wood_cavity_2014,degen_nuclear_2008}. This proposed resonant coupling for nuclear spins would be analogous to recent experiments and proposals utilizing direct resonant coupling between electron spins and 10 GHz-scale microwave cavities~\cite{bienfait_controlling_2016,haikka_proposal_2017}. 

In this article, we present work in which we harness a membrane resonator engineered for low mass and high-$Q$, known as a trampoline resonator~\cite{kleckner_optomechanical_2011,norte_mechanical_2016,reinhardt_ultralow-noise_2016} for spin detection~\cite{arcizet_single_2011,kolkowitz_coherent_2012,ovartchaiyapong_dynamic_2014}. Square SiN membranes have previously been used for MRFM~\cite{scozzaro_magnetic_2016}, torque magnetometry~\cite{Blankenhorn_membrane_2017}, and force-detected ESR~\cite{Takahashi_force_2018}. Here we follow the work of~\cite{scozzaro_magnetic_2016,Fedorov_Generalized_2019} by studying a significantly lower effective mass resonator, and develop a compact platform that will be conducive to cavity optomechanical integration. First, we demonstrate that trampoline resonators can maintain quality factors above 10$^6$ after being functionalized with a magnetic grain, and present a general framework for understanding force sensing using complex resonant modes. These complex modes allow effective separation of the optical and the mechanical interaction positions, essential for cavity integration. Second, as an initial spin-sensing demonstration, we detect an ensemble of fast-decaying electron spins of diphenylpicrylhydrazil (DPPH) at room temperature using a moderate magnetic field gradient created with a permanent magnetic grain affixed to the trampoline resonator. The detection scheme here is based on a Michelson interferometer, while an optical cavity integration is left for future work. We conclude with a discussion of the implications of our work for full cavity optomechanical integration of MRFM at cryogenic temperatures, and the detection of nuclear spins.

\section{Functionalization of high-tension mechanics}

\subsection{Concept}

Independent of the particular spin-sensing goal, a common thread in force-detected resonance is the effective functionalization of the mechanical element of choice.  To harness the high-$Q$ and complex geometries of Si$_3$N$_4$ resonators~\cite{tsaturyan_ultra-coherent_2017,Fedorov_Generalized_2019}, we must (1) learn to place magnets or samples on these resonators with minimal reduction of their quality factor and (2) understand through numerical modeling how mechanical geometry affects both the magnetic coupling and optical coupling as a function of spatial position on the resonator.

\begin{figure}[t]\centering \includegraphics[width=140mm]{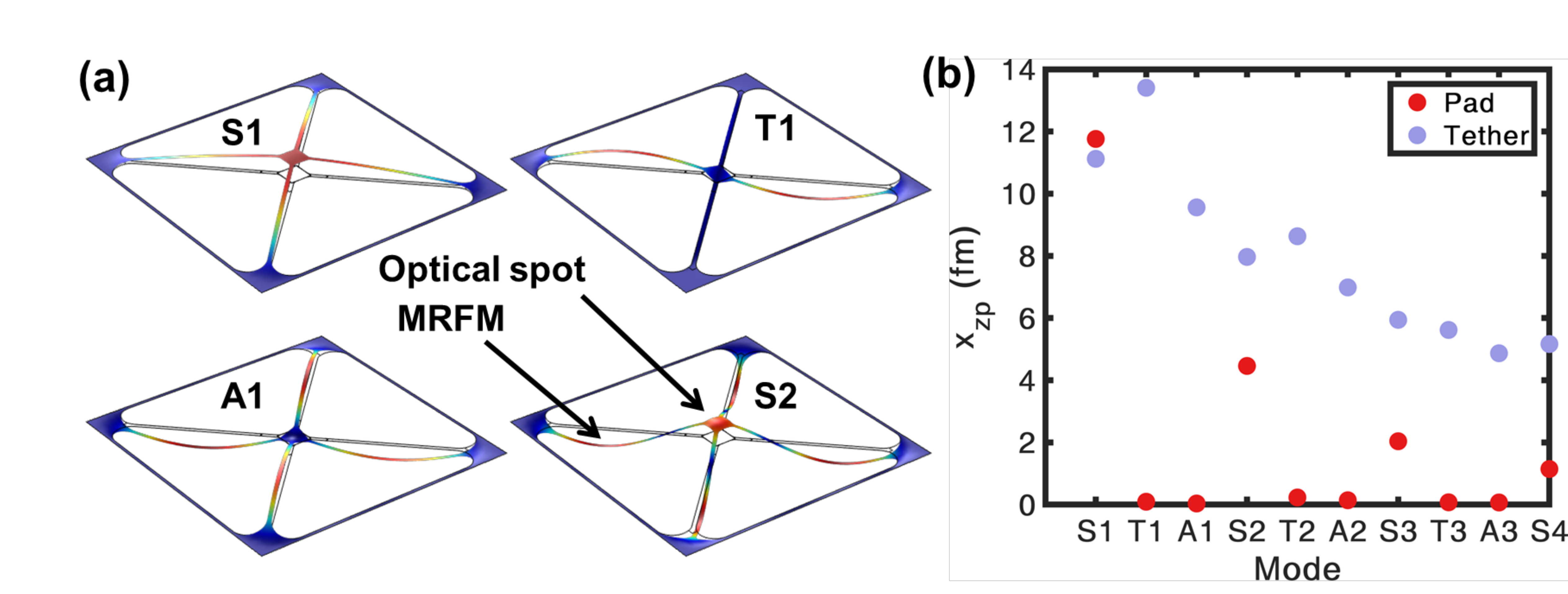}
\caption{Finite element simulation (COMSOL) of a \SI{500}{\micro\meter} wide and 30 nm thick trampoline with a pad size of \SI{30}{\micro\meter} and a tether width of \SI{2}{\micro\meter}. We note S1, T1, A1, S2 as the first symmetric, first torsional, first asymmetric and second symmetric modes, with frequencies of 359 kHz, 912 kHz, 936 kHz and 1140 kHz, respectively. (a) Trampoline mode shapes. Arrows indicate the cavity mode position and the magnetic coupling position, i.e.~the position where the magnetic grain or spin sample are deposited on the trampoline. (b) Simulated zero-point fluctuation of the trampoline pad (blue circles) and tether (red circles) for the first trampoline modes, up to S4, ordered from left to right by increasing resonant frequencies.}
\label{fig:modes}
\end{figure}

In our approach to functionalization, we spatially separate the optical cavity mode and the magnetic coupling achieved by depositing spins or a magnet on the mechanical resonator [Fig.~\ref{fig:modes}(a)]. To describe the associated interaction terms we find it convenient to use quantum scales; however this analysis translates well to the classical limit relevant to our proposed sensing applications. The optomechanical single-photon single-phonon coupling is $g_{0,\rm{OM}}=\hbar\frac{\omega_{\rm{cav}}}{L}\xzp$ \cite{aspelmeyer_cavity_2014}, where $\omega_{\rm{cav}}$ and $L$ are the cavity mode frequency, and the cavity length respectively, and $\xzp$ is the zero-point fluctuation. On the other hand, the spin-mechanical single-spin single-phonon coupling term (see \ref{HamiltonianAppendix}) is 
\begin{equation}
g_{\rm{SM}}=\hbar\gamma\frac{\partial B}{\partial x}\xzp.
\label{eq:g_SM}
\end{equation}
Here $\gamma$ and $\frac{\partial B}{\partial x}$ are the electron or nuclear gyromagnetic ratio and the magnetic gradient, respectively. Both coupling terms above are linear in $\xzp$; however, effectively, the zero-point fluctuation is position and mode dependent, due to the resonator spatial mode shape. The effective mass, $m_{\rm{eff}}$ that determines the zero-point fluctuation
\begin{equation}
\xzp=\sqrt[]{\frac{\hbar}{2 m_{\rm{eff}} \omega_m}}
\label{eq:x_zpf}
\end{equation}
is a function of the position and the mechanical mode:
\begin{equation}
m_{\rm{eff},j}(x,y)=m_{\rm{phys}}\frac{\int_{S_{\rm{tot}}} w_j^2(u,v)\rmd u\rmd v / \int_{S_{\rm{tot}}}\rmd u\rmd v}{w_j^2(x,y)},
\label{eq:m_eff}
\end{equation}
where $m_{\rm{phys}}$ is the physical mass of the resonator, and $w_j$ is the j-mode shape along the x-axis, $S_{\rm{tot}}$ is the total surface of the trampoline. Therefore, positioning the optical cavity mode and the magnetic grain at different locations allows optimization of the physical couplings. As an example in Fig.~\ref{fig:modes}(b), we show a simulation of $\xzp$ of several modes of trampoline, with a fundamental mode frequency of 359 kHz. $\xzp$ is calculated separately for the pad and the tether using Eqs.~(\ref{eq:x_zpf}, \ref{eq:m_eff}). We see that, for a given trampoline design, only the symmetric modes allow simultaneous opto-mechanical and spin-mechanical interaction. Moreover, the optimal MRFM detection position will depend both on the trampoline mode, as well as on the deposited sample.
Specifically, the sensitivity of magnetic force detection scheme is determined by the thermal vibration level of the resonator. The force noise power spectrum is given by the fluctuation-dissipation theorem:
\begin{equation}
S_F=\frac{4k_B T k}{\omega_m Q}=\frac{2\hbar k_B T}{Q \xzp^2},
\label{eq:s_f}
\end{equation}
where $k$ is the resonator spring constant, $Q$ is the quality factor, $\omega_m$ is the mechanical angular frequency. Therefore, to maximize force sensitivity, both the resonator's $Q$, along with its $\xzp$ should be maximized.

\subsection{Mechanical design and characterization of magnet-deposited membranes}

\begin{figure}[t]\centering \includegraphics[width=115mm]{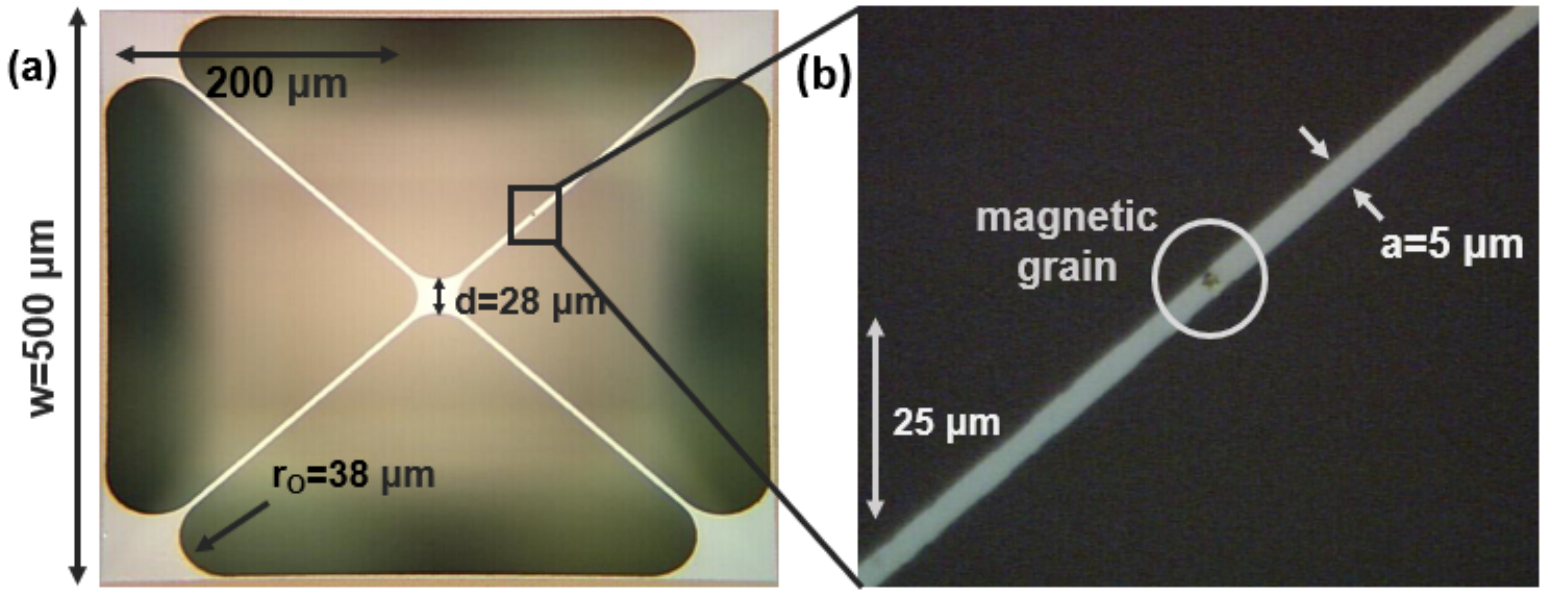}
\caption{ (a) SiN trampoline resonator functionalized with a NdFeB magnetic grain a few microns in diameter.  (b) Zoom in to magnetic grain deposited on resonator tether.}
\label{fig:components}
\end{figure}

We now present our experimental work on magnetic functionalization of trampoline resonators.  We use the 4-tether trampoline design of \cite{norte_mechanical_2016}, with the geometric parameters depicted in Fig.~\ref{fig:components}, and detailed in Table~\ref{table:tramp_dimen}, for two mechanical devices, marked A and B. The trampolines were deposited with an NdFeB magnetic grain with dimensions of a $2-3$ $\mu$m, to generate magnetic coupling between the trampoline and the spin sample. The grain was deposited on one of the tethers $\sim 100$ $\mu$m from the pad. Further details on the deposition procedure and the trampoline fabrication appear in \ref{AssemblyAppendix} and \ref{FabAppendix}, respectively.

The frequency and $Q$ of the mechanical resonators, before and after assembly, were measured by an optical setup based on a Michelson interferometer~\cite{barg_measuring_2018}. The resonators were mounted inside a vacuum chamber with a pressure of $\sim 10^{-6}$ torr. The interferometer signal beam was focused onto the trampoline pad with a spot size of \SI{30}{\micro\meter}. The mechanical resonant frequencies were identified within the device's thermal spectrum, while the quality factors were measured by resonantly exciting the modes by a ring piezoelectric actuator. The excitation is abruptly stopped and the energy decay time $\tau_m$ is extracted, where $Q=\omega_m\tau_m$. Resonant frequencies and $Q$ of two resonator chips before and after deposition appear in Table~\ref{table:deposition}. There, we see that although the epoxy and the magnetic grain are mechanically lossy, the small amount deposited still allow $Q$s above 10$^6$, while slightly reducing the resonance frequency. This optical setup was used for the magnetic force detection in this article as well, as described in the next section.

\begin{table}[ht]
\caption{\label{table:tramp_dimen} Critical dimensions of each trampolines in devices A and B. $t$ is the SiN thickness, $w$ is window size, $r_o$ is the outer fillet radius, $d$ the central pad length, $a$ is the tether width, and $m_{\rm{eff}}$ is the calculated resonator's effective mass of the fundamental mode, at the pad position, using Eq.~\ref{eq:m_eff}. Dimensions are depicted in Fig.~\ref{fig:components}.}

\begin{indented}
\item[]\begin{tabular}{@{}lll}
\br
Device & A & B \\
\mr
$t$ & 30 nm & 70 nm \\
$w$ & \SI{500}{\micro\meter} & \SI{500}{\micro\meter} \\                                                                               
$r_o$ & \SI{36}{\micro\meter} & \SI{15}{\micro\meter} \\                                                                                                 
$d$ & \SI{28}{\micro\meter} & \SI{28}{\micro\meter} \\                                                                                                                  
$a$ & \SI{5}{\micro\meter}  & \SI{5}{\micro\meter} \\ 
$m_{\rm{eff}}$ & 0.5 ng & 0.9 ng \\
\br
\end{tabular}
\end{indented}
\end{table}

\begin{table}[ht]
\caption{\label{table:deposition}Fundamental mode frequencies and quality factors of the resonators before and after deposition of the NdFeB magnetic grain, calculated spring constant and zero-point fluctuation (derived from the effective mass of Table ~\ref{table:tramp_dimen}, and the corresponding room-temperature force sensitivity, after deposition.}
\begin{indented}
\item[]\begin{tabular}{@{}llll}
\br
& Device & A & B \\
\mr
Trampoline & $f_{S1}$ & 429.1 kHz & 389.8 kHz \\
& Q & $4.5\times 10^6$ & $1.8\times 10^6$ \\
\mr
Trampoline with magnet & $f_{S1}$ & 413.2 kHz & 379.6 kHz \\
& Q & $2.4\times 10^6$ & $1.7\times 10^6$ \\
& k & 3.4 N/m & 5.3 N/m \\
& $\xzp$ & 6.3 fm & 4.9 fm \\
& $\sqrt{S_F}$ & 67 $\rm{aN}/\sqrt{\rm{Hz}}$ & 102 $\rm{aN}/\sqrt{\rm{Hz}}$ \\
\br
\end{tabular}
\end{indented}
\end{table}

\begin{figure}[t]\centering \includegraphics[width=100mm]{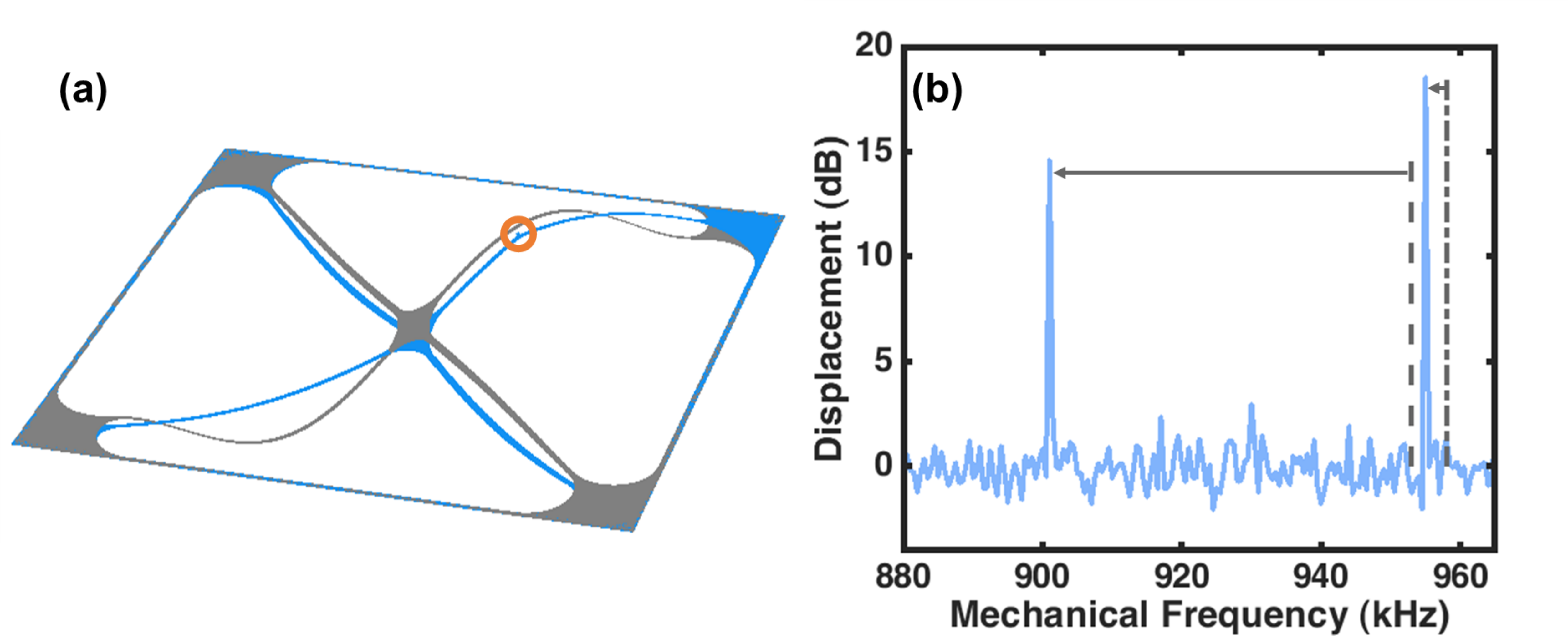}
\caption{Torsional-mode (T1) mode shapes and frequencies prior to (grey) and after (blue) deposition of a magnetic grain. (a) Simulated mode shapes of a free trampoline (grey) and a trampoline deposited with a cubic magnetic grain with an edge size of 2.5 $\mu$m (blue). The position of the deposited magnetic grain is marked by a red circle. (b) Measured trampoline displacement spectrum showing resonance location after deposition in blue. The original position of the resonances prior to deposition appear as vertical grey lines. The arrow show the reduction of the mode frequencies after deposition.}
\label{fig:torsional}
\end{figure}

After the magnetic grain is deposited, successful grain retention can be verified by the large frequency shift of one of the torsional T1 modes. For a perfectly symmetric bare trampoline, there are two degenerate T1 modes, representing the torsion of one pair of opposing tethers about the other perpendicular pair (Fig.~\ref{fig:modes}(a)). For Device A, the bare trampoline's slight asymmetry resulted in a few kHz shift between the two T1 modes (Fig.~\ref{fig:torsional} - grey). After deposition, however, the symmetry between the tethers is broken, and the frequency of the torsional mode associated with deposited grain decreases significantly, due to the added mass (Fig.~\ref{fig:torsional} - blue). This sensitivity to the presence of the grain can assist in estimating the mass of the deposited grain, as well as clearly indicating the integrity of the mechanical part.

\section{Electron-spin detection}

\subsection{Physical components}

\begin{figure}[t]\centering \includegraphics[width=120mm]{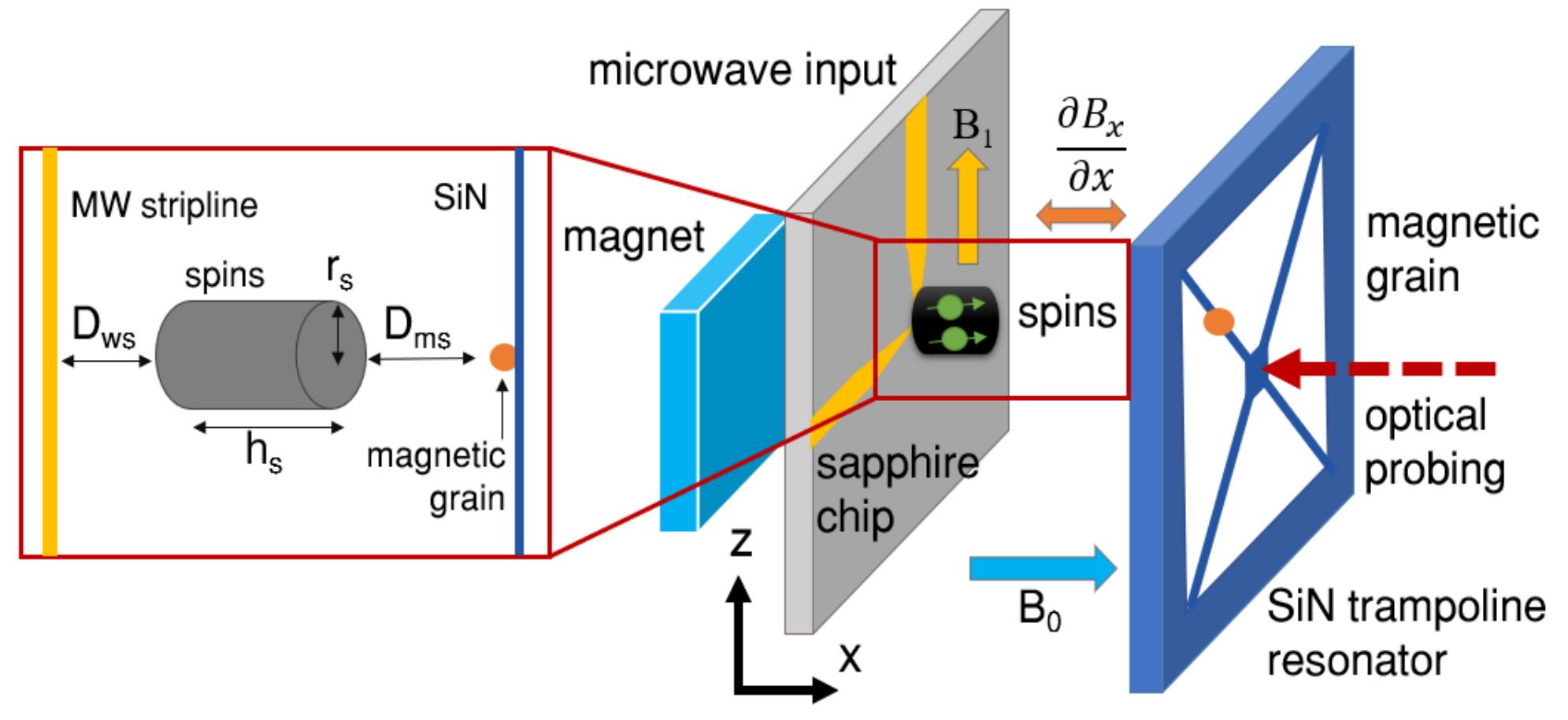}
\caption{Experimental schematic for magnetic resonance force microscopy demonstration.  We detect electron spins in DPPH with a trampoline resonator using cyclic saturation. Readout of the trampoline resonator is through a Michelson interferometer, with the signal arm reflected off either from the trampoline pad, or from a gold reflector on the sapphire chip.  Zoom in depicts device dimensions used in the simulation detailed in \ref{AnalysisAppendix}.}
\label{fig:setup}
\end{figure}

The apparatus we use for electron spin detection of DPPH is schematically depicted in Fig.~\ref{fig:setup}.  It is based on a two-chip design, which is comprised of a mechanical resonator chip deposited with a magnetic grain for magnetic gradient application, and sapphire chip with a deposited spin sample and a stripline for microwaves (MW) excitation of the spin resonance. The two chips are brought in close proximity to one another to enhance the magnetic force between the chips. A macroscopic permanent magnet sets the Larmor frequency of the spins. The displacement of the mechanical resonators is monitored by an optical Michelson interferometer. The two-chips and the magnets are placed in a vacuum chamber at room temperature and pumped to a vacuum level below $10^{-6}$ torr, to avoid gas damping of the mechanical resonator.

\subsection{Magnetic design}

To create the bias field $B_0$ we use a NdFeB permanent magnet with dimensions of $\frac{1}{2}"\times\frac{1}{2}"\times\frac{1}{8}"$ that sets a $B_0\sim 1000$ G magnetic field on the $x$-axis, parallel to the vibration direction of the mechanical resonator. $B_0$ is the dominant field that determines the spins' Larmor frequency $\omega_0=\gamma_e B_0$, where $\gamma_e=2.8/(2\pi)$ MHz/G for electronic spins. 

A magnetic grain of NdFeB, with a few \SI{}{\micro\meter} in dimensions was used for setting the magnetic gradient. The grain was magnetized with an electromagnet along the $x$-axis, and sets a gradient $\partial B_x / \partial x$. This gradient creates a force between the grain and the spins, equal to $F_x=M_x\cdot\partial B_x / \partial x$, where $M_x$ is the spin sample magnetization along the $x$-axis. The gradient along the $x$-axis is determined by the magnetic grain magnetization, size and shape, and its distance from the spin sample. The gradient along the magnetization axis of a spherical magnetic grain with remanent magnetization of $B_r$, radius of $R_M$ is
\begin{equation}
M_g=-\frac{2 R^3_M B_r}{r^4},
\label{eq:magnetization}
\end{equation}
where $r>R_M$ is the distance from the center of the grain. Typical dimensions for our setup included a grain with diameter of \SI{2}{\micro\meter}, at a distance of $10-20$ \SI{}{\micro\meter}, and remanent magnetization of 7 kG, which translates to a gradient between $0.1-1$ G/\SI{}{\micro\meter}. We note that the magnetization was calibrated by a magnetic property measurement system (MPMS) machine using 10 mg of magnetized NdFeB powder.  

\subsection{Electronic spins}
We utilize a DPPH spin sample of $20-30$ \SI{}{\micro\meter} in size. DPPH is used due to its high spin concentration of $\rho_{\rm{spin}} \approx 2.1\cdot 10^{21}$ spins/cm$^3$ and short relaxation time, of $25-80$ nsec, depending on the sample origin~\cite{Whitfield_Paramagnetic_1957, Goldsborough_Influence_1960, Schweiger_Pulsed_1988}. The DPPH thermal magnetization is $M_0=\frac{\chi_0 B_0}{\mu_0}$, where $\chi_0=2.5\cdot 10^{-5}$ is the DPPH magnetic susceptibility ~\cite{Itterbeek_static_1964}, $\mu_0$ is the vacuum permeability, and $B_0$ is the magnetic field. For relatively low magnetic gradients of $0.1-1$ G/\SI{}{\micro\meter} over the DPPH grain, we can estimate the number of spins contributing to the force by simply $V_{\rm{DPPH}}\cdot\rho_{\rm{spin}}\approx 10^{13}$.

\subsection{Microwave application}

A gold stripline deposited on a sapphire substrate delivers a MW tone that drives spin manipulation, as seen in Fig.~\ref{fig:apparatus}. A sapphire substrate is chosen due to its relatively high thermal conductivity and high electrical resistivity. The stripline is designed as a 90$^{\circ}$ corner for application of the MW field, with $B_1 \ll B_0$, perpendicular to $B_0$ in the y-z plane. The narrowest section of the corner reaches $\sim$ \SI{20}{\micro\meter}, on the same order of the DPPH grain. The MW is generated by a signal generator and a 3 W amplifier. For 5 mA of current ($\approx1$ mW), we estimate $B_1\approx 0.5$ G $\Rightarrow \Omega_{\rm{Rabi}}\approx 2\pi\times 1.5$ MHz. However, experimental results suggest high loss in the MW delivery, resulting in $B_1\approx 50$ mG. More details can be found in \ref{AnalysisAppendix}.

\subsection{MRFM detection}
\label{secMRFM}

We use an MRFM detection technique known as cyclic saturation~\cite{scozzaro_magnetic_2016, wago_magnetic_1997,Rugar_mechanical_1992}. This detection method is appropriate if the spin relaxation time is much shorter than the resonator's period $\tau\ll T=\frac{2\pi}{\omega_m}$. In this case, if a MW tone is modulated at a frequency resonant with one of the mechanical modes the steady-state spin magnetization can be expressed, according to the Bloch equations \cite{wago_magnetic_1997}. We assume $\tau=T_1=T_2$, which is typical for DPPH ~\cite{Whitfield_Paramagnetic_1957, scozzaro_magnetic_2016}:
\begin{equation}
M_x=M_0\bigg[1-\frac{\gamma^2 B_1^2\tau^2}{1+(\gamma B_0-\omega_{\rm{MW}})^2\tau^2+\gamma^2 B_1^2\tau^2}\bigg],
\label{eq:Mz}
\end{equation}
where $\omega_{\rm{MW}}$ is the MW angular frequency, and $B_1$ and $B_0$ are the position dependent RF and DC magnetic fields respectively. We utilize two types of cyclic saturation schemes - amplitude modulation (AM), where the amplitude of $B_1$ is modulated $A_m(t)=A_0\cdot(1+m\cdot \sin(\omega_m t))$, and frequency modulation (FM), where its frequency is modulated $\omega_{MW}(t)=\omega_{0}+\omega_{FM}\cdot \sin(\omega_m t)$, with $\omega_{FM}$ the frequency deviation. To calculate the magnetic force applied on the resonator we integrate over the magnetic field distribution, assuming radial symmetry with a magnetic grain magnetized along the symmetry axis:
\begin{equation}
F=2\pi\int_{x_{min}}^{x_{max}}\int_0^{r_{max}(x)} M_{1}(x,r)\cdot r\cdot\frac{\partial B_x}{\partial x}\bigg|_{x,r} \rmd r \rmd x,
\label{eq:ForceAM}
\end{equation}
where $x_{min}-x_{max}$, $r_{max}(x)$ are the DPPH grain boundaries on the x-axis and r-axis, respectively. $M_1$ is the Fourier component of the modulation at an angular frequency of $\omega_m$. In the case of AM (full modulation) $M_{1,\rm{AM}}=\frac{2}{\pi}(M_0 - M_x)$, and for FM $M_{1,\rm{FM}}=\omega_{\rm{FM}}\cdot\frac{\partial M_x}{\partial \omega_{\rm{MW}}}$ ~\cite{wago_magnetic_1997}. We note here that detection based on FM is more robust versus electrical spurious forces compared to AM, and therefore the main results in section \ref{results} are measured in FM. 

We use two methods to sweep over the magnetic resonance. The first is varying the MW frequency with a fixed magnetic field, and the second is fixing the MW frequency while varying $B_0$. To sweep $B_0$ we add a second NdFeB magnet with dimensions of $2 '' \times 1 '' \times \frac{1}{2} ''$ outside the vacuum chamber, and varied the field between $25-43$ G, with an opposite magnetization to the first magnet. These fields correspond to a distance of 87 to 75 mm from the flip-chip. The second magnet was added to enable fine scan of $B_0$, with a mm-resolution displacement, outside of the vacuum chamber.

A complication in the detection is a slow drift of the resonator's frequency. The undamped trampoline resonators used here had a linewidth of $\sim 0.2$ Hz, with a few Hz drifts at time scales of a tens of seconds. Future experiments will use passive damping provided by a cavity, but for these initial experiments we simply avoid this drift by sweeping over the mechanical resonance with a range of $10-30$ Hz. We verified that the driven amplitude reaches its steady-state value, when increasing the sweep duration. As noted in the discussion section, damping of the resonator would reduce or eliminate the impact of the drifts for coherent detection performance as well as for the efficiency of the acquisition sequence.

\begin{table}[ht]
\caption{\label{table:flipchip_dimen} Critical dimensions of each flip chip were first estimated from optical microscopy and finite element simulations. Further refinements were determined by fitting data to the model provided by Eq. \ref{eq:ForceAM}. $r_m$ is the radius of the magnetic grain, $D_{ms}$ is the distance between magnetic grain and spin ensemble, $D_{sw}$ is the distance between spin grain and MW stripline, $r_s$ is the radius of spin ensemble (cylinder), $h_s$ is the height of spin ensemble (cylinder) $B_1$ is the MW-frequency magnetic field, $\tau$ is the relaxation time of spins, $B_0$ is the uniform magnetic field, and $\frac{\delta B_0}{\delta x}$ is the local gradient of uniform magnetic field.}

\begin{indented}
\item[]\begin{tabular}{@{}lll}
\br
Device & A & B \\ 
\mr
$r_m$ & \SI{1.5}{\micro\meter} & \SI{1.45}{\micro\meter} \\ 
$D_{ms}$ & \SI{5}{\micro\meter} & \SI{8}{\micro\meter} \\
$D_{sw}$ & \SI{5}{\micro\meter} & \SI{5}{\micro\meter} \\
$r_s$ & \SI{15}{\micro\meter} & \SI{15}{\micro\meter} \\
$h_s$ & \SI{35}{\micro\meter} & \SI{25}{\micro\meter} \\
$B_1$ & 50 mG & 80 mG \\
$\tau$ & 42 ns & 42 ns \\
$B_0$ & 915.7 G & 873.8 G \\
$\delta B_0/\delta x$ & 20 G/mm & 25 G/mm \\
\br
\end{tabular}
\end{indented}
\end{table}

\subsection{MRFM results and analysis} \label{results}

\begin{figure}[t]\centering \includegraphics[width=160mm]{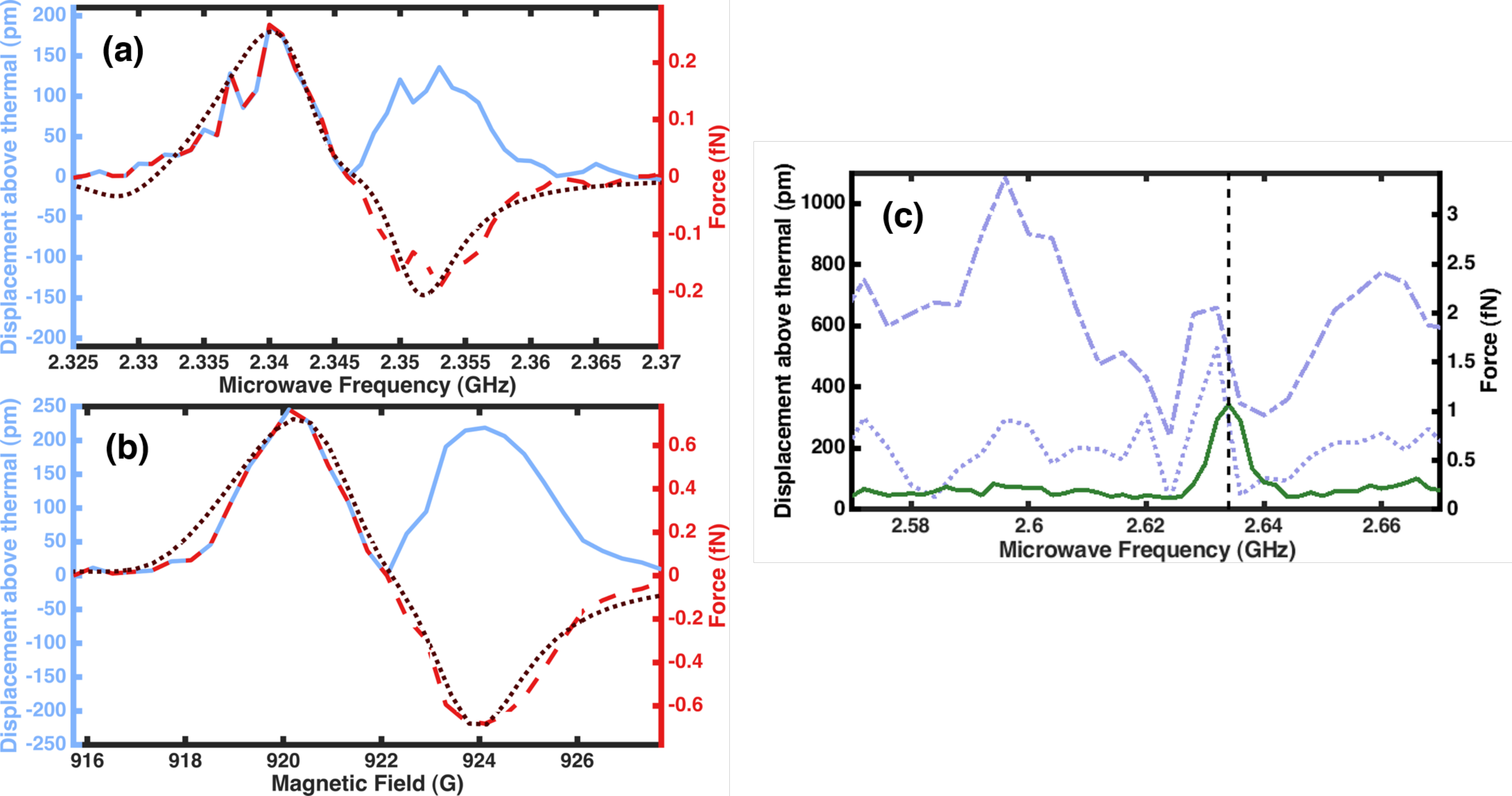}
\caption{(a) MRFM resonance from device A using an FM microwave drive. The MW frequency is swept at a fixed magnetic field. Shown are the mechanical displacement (full blue), the corresponding force signal (dashed red), and a fit of the FM signal (dotted black). (b) Same as (a), but here we sweep the magnetic field at a fixed microwave frequency of 2.564 GHz and use device B. In both (a) and (b), the input microwave drive power is -7 dBm. (c) MRFM resonance using an AM microwave drive of device B.  We show a sweep of the MW frequency at a fixed magnetic field at three different MW powers of -8 dBm (full green), -3 dBm (dotted blue), and 0 dBm (dashed blue).  Particularly using the AM technique, the spin-resonance signal can easily be overwhelmed by spurious electrical forces, as observed at the higher powers (dotted blue, dashed blue).}
\label{fig:FMScan}
\end{figure}

Figure ~\ref{fig:FMScan}(a) shows a MW frequency sweep of device A, with FM cyclic saturation. The sweep parameters are $\omega_{\rm{FM}}=2\pi \times 10$ MHz, $B_1\approx 50$ mG. A sweep of the $B_0$ magnitude with FM cyclic saturation of device B appear in Fig.~\ref{fig:FMScan}(b). There, the MW frequency was $\omega_{\rm{MW}}=2\pi\times 2.564$ GHz, and $B_1 \approx 100$ mG with modulation deviation of $\omega_{\rm{FM}}=2\pi\times 10$ MHz. The displacement amplitude was calculated either from the calibrated Michelson setup [Fig. ~\ref{fig:FMScan}(a)] (device A) or by comparing the displacement enhancement over the estimated thermal signal [Fig. ~\ref{fig:FMScan}(b), ~\ref{fig:FMScan}(c)] (device B). For a calibrated Michelson interferometer $A=V_p\cdot C$, where $C$ is the Michelson calibration factor in V/nm, and $V_p$ is the Michelson amplitude in volts at the mechanical resonance frequency. By measuring the enhancement $E$ over the thermal signal $x_{th}=\sqrt{k_B T / k}$, we extract the signal amplitude of $A=x_{th}\cdot E$. To estimate the applied force we calculate the trampoline response to a steady-state resonant excitation as $F=(A\cdot k)/Q$. The conversion factor of $k/Q$ was different between the two chips, as seen in Table \ref{table:deposition}. We note that in all of the measurements only the signal amplitude was acquired (blue), and a sign change was added to the FM force analysis to accommodate the expected phase flip of the FM resonance (red dashed). To estimate the geometric and physical parameters of the flip-chip, a fit to the 2D-magnetic force model, using Eq.~\ref{eq:ForceAM}, was performed, as seen in Fig.~\ref{fig:FMScan}(a) and in Fig.~\ref{fig:FMScan}(b). From the fit we derive the flip-chip parameters as appear in Table \ref{table:flipchip_dimen}. We note that the parameter values were within the estimated range. More details of the MRFM fitting procedure for derivation of the values in Table \ref{table:flipchip_dimen} appear in  \ref{AnalysisAppendix}.

The main limitation for increasing $B_1$ was the electric spurious force between the two chips. Spurious forces are typical for MRFM setups. However, with this initial flip-chip design we did not apply the various methods intended for coping with electric spurious forces, such as those described in ~\cite{Poggio_Nuclear_2007, nichol_nanomechanical_2012, Longenecker_HighGradient_2012, scozzaro_magnetic_2016}. Unlike the magnetic force, the spurious force is broadband yet structured, and therefore interferes with the magnetic resonance, and at high MW powers overwhelms it.  The spurious force is generally more apparent in AM spectra, as we see in (Fig.~\ref{fig:FMScan}(c)).  In future design and experimental sequence one must consider mitigation of the spurious force to allow more efficient resonant magnetic drive.

\begin{figure}[t]\centering \includegraphics[width=100mm]{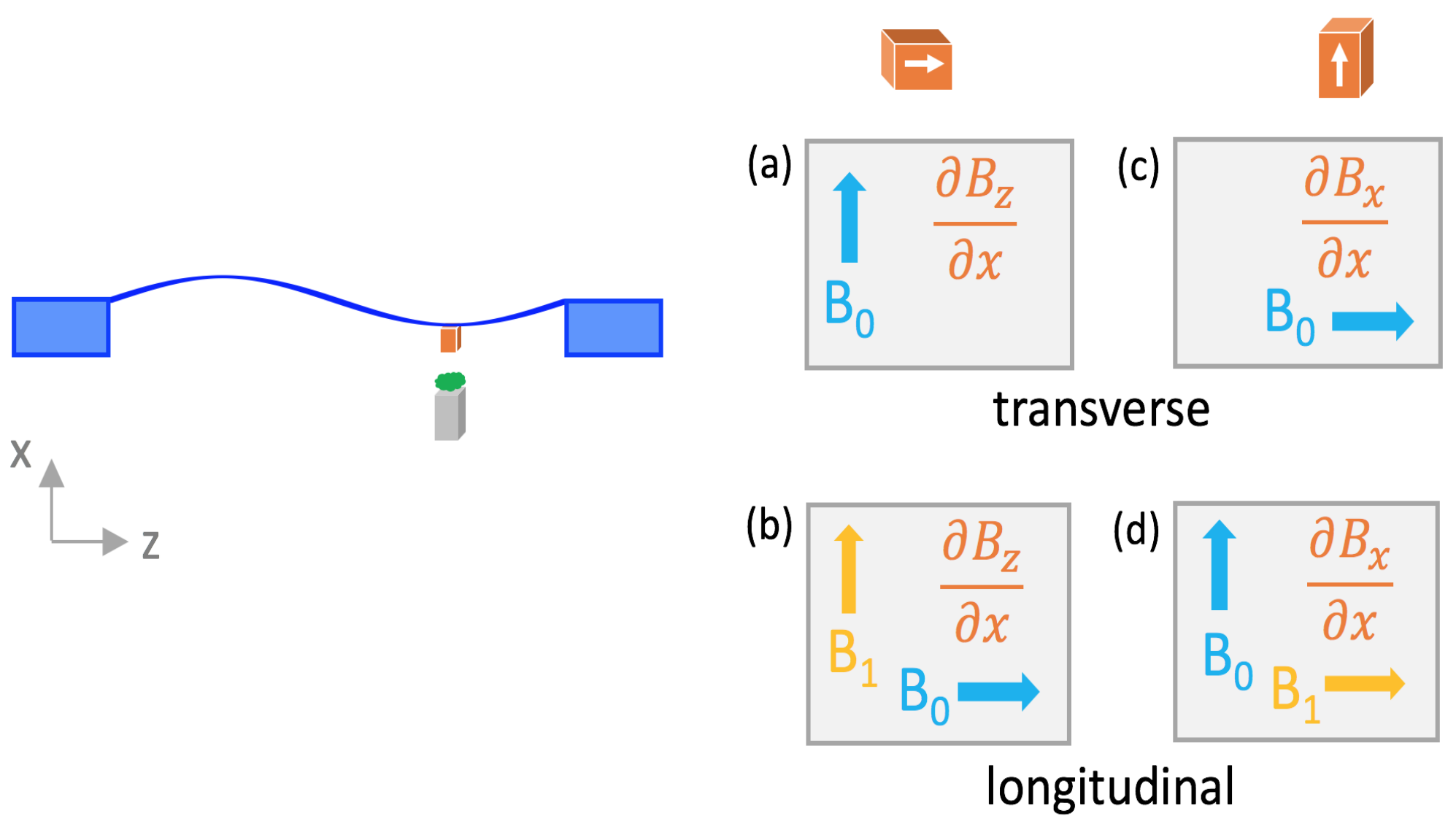}
\caption{Possible magnetic field configurations of $B_0$, $B_1$ and the magnetic gradient for transverse detection [(a), (c)] and for longitudinal detection [(b), (d)]. The magnetic grain is represented by the orange box with the white arrow indicating its magnetization direction.  One envisions the magnetic fields as being created in the context of the geometry of Fig.~\ref{fig:concept}, and the choice of configuration dependent on the type of coupling or detection required.  For the DPPH detection presented in Sec.~\ref{secMRFM}  we use configuration (d).  For future resonant detection one would choose (a) or (c).}
\label{fig:bdirection}
\end{figure}

\section{Discussion}

In the experiments above we laid the groundwork for an integrated platform that couples spins to engineered membrane resonators.  We now discuss the prospect for bringing together state-of-the-art tensioned mechanics, magnetic couplings, and high-finesse cavities, as schematically depicted in Fig.~\ref{fig:concept}. 

\subsection{Cavity and mechanical integration}

Efficient damping is a key component of high-bandwidth spin detection~\cite{Poggio_feedback_2007}. In our initial work we simply used an undamped trampoline resonator; however, the relaxation time of a high-$Q$ resonator with frequencies of $1-10$ MHz can exceed seconds, and even minutes, which is prohibitive to detection protocols.  Optical damping facilitated by a high finesse cavity via optomechanical interaction provides a natural method that incorporates high displacement sensitivity and that does not require high-frequency feedback loops (Fig.~\ref{fig:concept})~\cite{aspelmeyer_cavity_2014,nichol_nanomechanical_2012}. 

Fabry-Perot cavities with finesse of a few thousand are routinely integrated with membrane resonators~\cite{thompson_strong_2008,purdy_observation_2013,peterson_laser_2016,rossi_measurement_2018}, and damping rates upward of 10 kHz have been demonstrated in a cryogenic environment down to 100 mK~\cite{peterson_laser_2016}.  Note that resonator designs with $\xzp\sim 10$ fm, and $\omega_m\leq 2\pi\times 10$ MHz imply structures of $\sim$ \SI{10}{\micro\meter} in size.  Hence, experiments may require cavities with small mode sizes, which can be achieved with fiber cavities~\cite{Flowers_Fiber_2012} or integrated silicon mirrors~\cite{kuhn_nanoparticle_2017}.  Integrated photonic crystals cavities or whispering gallery mode cavities in conjunction with a variety of high-sensitivity mechanical resonators offer additional prospects~\cite{schliesser_high-sensitivity_2008,eichenfield_optomechanical_2009}.
 
Further, future prospects of this platform will require increasing mechanical quality factors from the $10^6$ level demonstrated here, to the $10^8$ level required for projected numbers, while achieving both optical and magnetic integration. In this article, we demonstrated that when depositing a micron-size grain on a trampoline tether, a $Q>10^6$ could be maintained, along with $\xzp$ of $5-7$ fm. To utilize even higher $Q$ devices of $10^8$ or higher ~\cite{tsaturyan_ultra-coherent_2017, Ghadimi_elastic_2018}, a sample (either a magnetic grain or a spin sample, as in Fig.~\ref{fig:concept}) with smaller dimensions should be deposited. We expect that deposition of magnetic sample with dimensions of less than a micron will require a more integrated deposition method. 

\subsection{Magnetic resonance}

The strength of the magnetic gradient determines the magnetic force magnitude ($F=\hbar \gamma G$, for a single spin) as well as the coherent coupling strength $g_{\rm{SM}}$. Both are crucial for increasing spin sensitivity and resolution, and note that ideas such as spin cooling through mechanics requires extreme couplings. In this work, we demonstrated moderate gradients of up to 1 G/\SI{}{\micro\meter}, whereas record gradients reached 60 G/nm ~\cite{Mamin_Gradient_2012}, almost 5 orders of magnitude stronger. The main challenge in designing a force sensor based on stressed membranes or doubly-clamped strings is the ability to create a separation of tens of nm between the gradient source and the spins that will allow gradients above 10 G/nm. One can envision a scanning stiff pillar with a gradient source at its end, as depicted in Fig.~\ref{fig:concept} that will allow relative position scanning at distances of 100 nm or below. 

Application of a strong enough $B_1$ that will allow rapid manipulation of the spins is also an essential component for most MRFM applications. The flip-chip design applied here, with MW application by a non-resonant stripline, achieved only limited $B_1$ values, below 0.2 G. The stripline design, common in MRFM, is usually integrated with the magnetic gradient source as well~\cite{degen_nanoscale_2009, nichol_nanomechanical_2012}. The inherent 2D geometry of tensioned resonator limits stripline designs and a different approach such as a resonant MW circuit ~\cite{scozzaro_magnetic_2016} should be explored.

Lastly, we review some considerations for spin-mechanics resonant coupling, i.e.~operating in a domain where the mechanical resonance and the spin Larmor frequency are matched; magnetic resonance imaging experiments have to-date operated with cantilevers with resonant frequencies much smaller than spin frequencies.  The associated Larmor frequencies for fields above 3000 G are greater than $10$ GHz for electron spins and greater than $10$ MHz for nuclear spins. This implies that tensioned SiN devices are suitable for resonant coupling with nuclear spins. (Resonant coupling with electrons could potentially be achieved with silicon optomechanical crystal resonators ~\cite{chan_laser_2011,meenehan_silicon_2014}, and has been achieved with microwave electromagnetic resonators~\cite{bienfait_controlling_2016}.)

Any detection scheme harnessing resonant detection will be inherently different than standard longitudinal MRFM detection. Because the mechanical and the Larmor frequencies are equal, manipulating the spins at that rate would require strong $B_1$ fields, as well as breaking the rotating wave approximation. This would make the spin control very challenging. Therefore, two detection schemes can be considered. The first is detecting the spins by starting with a $\pi/2$ pulse followed by Larmor precession that would resonantly drive the spin oscillator. The second is detecting the normal mode splitting or broadening due to strong coupling between the two quantum systems. Both detection schemes require a magnetic gradient of the perpendicular field, as shown in geometries (a),(c) in Fig.~\ref{fig:bdirection}. Meaning, for $B_0\hat{x}$, the gradient would be along $\frac{\partial B_z}{\partial x}$. This is unlike the parallel gradient direction of the standard MRFM, aimed at detection of longitudinal magnetization, which appear in realizations (b),(d) in Fig.~\ref{fig:bdirection}. 

\section{Conclusion}

In this work we explored multiple facets of using engineered SiN tensioned membranes for magnetic force sensing.  We separated the optical and magnetic interaction positions by depositing a magnetic grain on the trampoline tether and optically measuring the central pad. We integrated these resonators in a flip-chip design and sensed an ensemble of electronic DPPH spins at a force level of 0.1 fN, which improves force sensitivity by almost 2 orders of magnitude over previous spin sensing with tensioned resonators ~\cite{scozzaro_magnetic_2016}.  Our explorations will instruct future integration with a high-finesse cavity for implementation of optomechanical concepts.  We discussed future prospects for using tensioned membrane resonators for resonant interaction with nuclear spins, where we envision a major challenge is achieving gradients above 10 G/nm with planar tensioned resonators.

\ack

We would like to thank Dan Rugar, John Mamin, Brad Moores, Martino Poggio, Christian Degen, Richard Norte, Darrick Chang, Thomas Purdy, and Peter Rabl for fruitful discussions, Joshua Biller and Gareth Eaton for spin physics discussions and assistance with EPR measurements, Stephen Russek for assistance in magnetic MPMS measurements, Brett Fiedler and Ralph Jimenez for optical microscopy assistance, and John Teufel for critical manuscript reading. We acknowledge funding from AFOSR PECASE, the NSF under grant number PHYS 1734006, the CU UROP program, and a Cottrell Scholar award.

\appendix

\section{Spin-mechanics Hamiltonian} \label{HamiltonianAppendix}

The force between a magnetic moment $\vec{\mu}$ and a gradient source $\frac{\partial \vec{B}}{\partial x}$ on a mechanical resonator vibrating on the x-axis (Fig. \ref{fig:bdirection}) is $F_x= \vec{\mu} \cdot \frac{\partial \vec{B}}{\partial x}$. For a single spin (either for electronic or nuclear spin) $\vec{\mu}=\mu_{\rm{B,N}}\cdot \vec{S}$. Then, the spin-mechanics Hamiltonian can be formulated as $H_{\rm{SM}}=g F_x \cdot x$, where $g$ is the spin $g$-factor. We express $\vec{S}$ and $x=\xzp \cdot (b+b^\dagger)$ as operators, where $b^\dagger$, $b$ are the phonon-states raising and lowering operators, and formally derive
\begin{equation}
H_{\rm{SM}}=\hbar \gamma \xzp (b+b^\dagger)\frac{\partial \vec{B}}{\partial x}\cdot \vec{S},
\label{eq:SM_hamiltonian}
\end{equation}
where $\gamma=g \mu_{\rm{B,N}} / \hbar$ is the gyromagnetic ratio. Therefore, we derive the coupling term of Eq.~~\ref{eq:g_SM} as the pre-factor of $H_{\rm{SM}}$, in the presence of a magnetic gradient $\frac{\partial \vec{B}}{\partial x}$ as $g_{\rm{SM}}=\hbar \gamma \xzp\frac{\partial \vec{B}}{\partial x}$.

\section{MRFM signal analysis} \label{AnalysisAppendix}
Fits to the FM modulated cyclic-saturation signals were achieved by numerically calculating Eq.~\ref{eq:ForceAM} with $M_1(x,r) = \omega_{\rm{FM}}\cdot\frac{\partial M_x}{\partial\omega_{\rm{MW}}}$. Initial parameter values were determined from estimated and literature values and were varied within reasonable uncertainties until a qualitative correspondence between the simulated and measured lineshape was reached. The lineshape of the MRFM signal has a strong dependence on the device geometry outlined in Fig.~\ref{fig:setup}. The spatial extent of the spin sample gives rise to an inhomogeneous lineshape determined by spatial dependence of $B_1$, $B_0$, and $\frac{\partial B_x}{\partial x}$ given by (notations are according to Table \ref{table:flipchip_dimen}):

\begin{equation}
B_1(x) = \frac{\mu_0 I}{2 \pi (h_{s}- x + D_{sw})}
\label{eq:B1}
\end{equation}
\begin{equation}
\vec{B}_0(x,r) = \vec{B}_{m}(x,r) + \vec{B}_{dip}(x,r)
\end{equation}
\begin{equation}
\vec{B}_m(x,r) = (B_0 + \frac{\delta B_0}{\delta x}x)\hat{x}
\end{equation}
\begin{equation}
\frac{\partial B_x}{\partial x}(x,r) = \frac{\partial (B_{dip})_x}{\partial x}\bigg|_{x,r}
\end{equation}

where $\vec{B}_{dip}(x,r)$ is a magnetic dipole field originating from the center of a spherical magnetic grain with remnant magnetization $B_r$ (Eq.~\ref{eq:magnetization}), and $\vec{B}_m(x,r)$ is the field from the large permanent magnet illustrated in light blue in Fig. \ref{fig:setup}. The integration region of the spin is taken to be a cylinder such that $r_{max}(x)= r_{s}$. The $B_1$ field is modeled as the field from a thin wire elbow carrying a current $I$. $B_{0}$ determines the Larmor frequency of the magnetic resonance, while $\frac{\delta B_0}{\delta x}$ corresponds to the spatial gradient of magnetic field from the large permanent magnetic. This gradient is extracted from a finite element simulation (COMSOL) of the large magnet. The fits for the observed MRFM signals indicate that the observed $B_1$ is significantly weaker than the calculated field from the expected MW transmission; this is consistent with 20 dB loss in the microwave line after the microwave amplifier. 

\section{Flip-chip assembly} \label{AssemblyAppendix}

The flip-chip device is assembled in three steps: First, the magnetic grain and the DPPH grain are deposited on the resonator and sapphire chips, respectively. Both grains are attached using G1 epoxy from Gatan Inc. Glass tips are used for epoxy and grain deposition, and are maneuvered with a 3-axis micropositioner.  Second, after the epoxy has cured, the resonator chip is placed between the poles of an electromagnet to magnetize the NdFeB magnetic grain. Third, the two chips are positioned opposite to one another at a distance of $1-10$ \SI{}{\micro\meter} between them, and secured with stycast epoxy, deposited with a glass tip. The sapphire bottom chip of device B has a gold reflector opposite to the trampoline pad to enhance the optical reflection, while device A has a clear optical path for no reflection.

\begin{figure}[t]\centering 
\includegraphics[width=50mm]{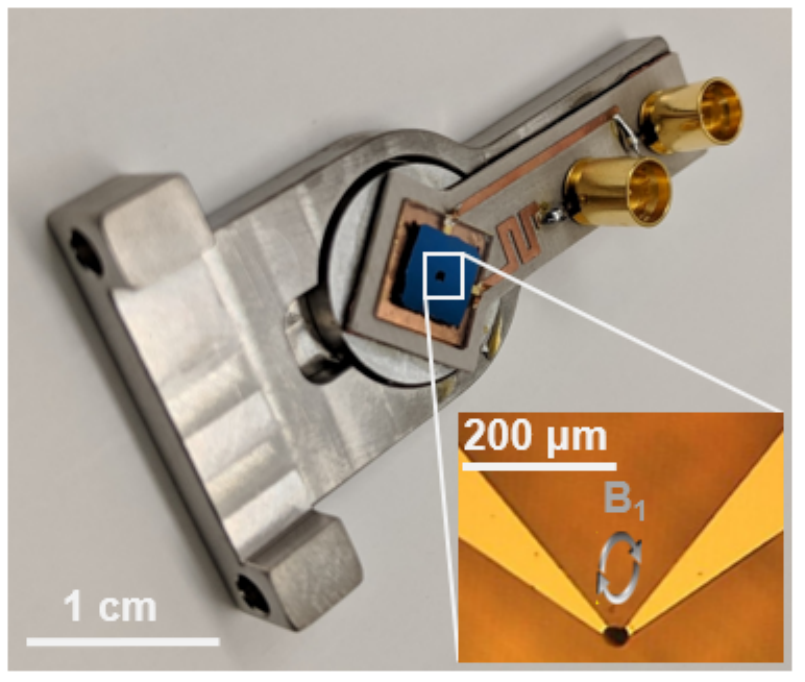}

\caption{Flip chip device connected to a MW PCB, and positioned on a titanium holder. Inset shows MW stripline design with the deposited spin sample and a reflector.}
\label{fig:apparatus}
\end{figure}

\section{Trampoline fabrication} \label{FabAppendix}

Resonators are fabricated from a stoichiometric silicon-nitride (Si$_3$N$_4$) clad (80 or 40 nm) silicon wafer (\SI{385}{\micro\meter}). The wafer is diced into 5 mm square chips and patterned on both sides. The resonator is written into a PMMA photoresist using a scanning electron microscope. The opposite side of the chip is patterned with a positive UV resist. The exposed Si$_3$N$_4$ is etched with a sulfur-hexaflouride reactive ion etch; the silicon is then etched from both sides in a 30\% KOH solution for 3.5 hours to release the trampoline.  The chip is then cleaned in Nanostrip.

\section*{References}
\providecommand{\newblock}{}

\end{document}